\documentclass[prl,a4paper,twocolumn,showpacs,superscriptaddress,amssymb]{revtex4}
\usepackage{graphicx}

\begin {document} 
%\draft %Obsolite in revtex4

\title{Frequency Dependence of the Critical Velocity of a Sphere Oscillating in Superfluid Helium-4}
\author{R. H\"anninen}
\affiliation{Low Temperature Laboratory, Helsinki University of Technology, FIN-02015 TKK, Finland} %Revtex4
%\address{Low Temperature Laboratory, Helsinki University of Technology, FIN-02015 TKK, Finland} %Old revtex
\author{W. Schoepe}
\affiliation{Institut f\"ur Experimentelle und Angewandte Physik, Universit\"at Regensburg, D-93040 Regensburg, Germany} %Revtex4
%\address{Institut f\"ur Experimentelle und Angewandte Physik, Universit\"at Regensburg, D-93040 Regensburg, Germany} %Old revtex
\date{January 16, 2008}

%\maketitle %must be used after abstract when using new revtex4 format

\begin {abstract}
It is shown that the critical velocity of a small sphere oscillating in superfluid helium increases with the square root of the oscillation frequency. This behavior can be described by a simple dimensional argument. The size of the sphere and the temperature of the superfluid are found to have no or only very little effect. Surface properties of the sphere and remanent vorticity may have an influence but have not been under systematic investigation in these measurements.  
\end{abstract}


\maketitle %New revtex4 format

%No multicols needed for revtex4
%\begin{multicols*}{2}  % ###multicol###
A quantitative description of the critical velocity for the onset of turbulent drag of a macroscopic body moving in superfluid helium is still an unsolved problem although there is a large amount of experimental data available from vibrating wires, grids, quartz tuning forks, and oscillating spheres. At small driving forces $F$ the velocity amplitude $v$ grows  linearly because the drag force due to quasiparticle scattering is proportional to the velocity. At a certain critical value $v_c$ vorticity is shed by the moving body and the drag increases strongly. Typically, these velocity amplitudes are in the range from mm/s up to almost m/s \cite{Vinen}, and therefore much lower than the Landau critical velocity of about 50 m/s. Above $v_c$ the velocity amplitude grows approximately with the square root of the driving force because turbulent drag is proportional to the square of the velocity \cite{PRL1}. Critical velocities for vortex production have been calculated, for a review see \cite{Donnelly}. For example, a vortex ring of radius $R$ is calculated to expand if a critical velocity is exceeded that is given approximately by \cite{Feynman}

\begin{equation}
v_c=\frac{\kappa}{2\pi R}\cdot \ln \left(\frac{8R}{a_0}\right),
\label{e.critvel} 
\end{equation}
where $\kappa \approx 10^{-7} $m$^2$/s is the circulation quantum and $a_0\approx 0.15$ nm is the size of the vortex core or the coherence length. For a sphere of radius $R$ = 100 $\mu$m this would give 2.5 mm/s for a ring of the same size. This is at least an order of magnitude lower than the experimental value, see below. Within this picture, it follows that smaller vortex rings will be shed, probably because of the roughness of the surface. Numerical simulations of vortex shedding have been performed using full Biot-Savart calculations and assuming a smooth sphere \cite{Vinen,movie}. Numerically, the obtained critical velocities are of same order as in experiments. However, the results are somewhat sensitive to the computational accuracy and therefore the frequency dependence for $v_c$ is not extracted yet. In any case, a simple dimensional argument tells us that a characteristic velocity scale in the turbulent superfluid must be given by $\kappa/R$, whatever the length scale R is.

The question arises whether the physics will be different in case of oscillatory motion. The turbulent drag force $F_D=\gamma v^2-F_0  \; (\gamma = C_D \rho \pi R^2/2$, where $\rho$ is now the superfluid density and $C_D\approx0.4$ for a sphere \cite{Landau}) differs from the classical result for steady flow only by the shift $F_0$ \cite{PRL1} .
This shift implies a critical velocity for the onset of turbulent drag given by $v_c=\sqrt{F_0/\gamma}$. The shift is temperature independent \cite{OVL}, and so is this critical velocity. A theoretical calculation of $F_0$ is not available. (It should be mentioned, however, that a similar drag force has been calculated for a body moving through a Bose condensate \cite{Adams}.) In this situation it is necessary to resort to dimensional considerations.
Because there is a time scale now given by the oscillation frequency $\omega$, we have a new velocity scale  
for the onset of the turbulent regime given by \cite{risto} 

\begin{equation}
v_c\sim\sqrt{\kappa \cdot \omega}\label{e.scalerel}. 
\end{equation}
Remarkably, there is no length scale present any more. There might be, however, a logarithmic term like the one in Eq.(1). This implies that this critical velocity does not or only very little depend on the size of the body. In the following, this scaling law is confirmed experimentally with spheres oscillating at various frequencies. In Appendix 1 a model calculation based on Kelvin waves is presented that supports the scaling law. In the Appendix 2, the same dimensional arguments are applied to a classical viscous liquid and shown to be consistent with analytical calculations for spheres.

The experiments were performed at Regensburg University in the years from 1993 to 2000. Most of the results have been published in earlier work \cite{all papers}, but a systematic analysis of the frequency dependence of the critical velocities was only performed recently. The data were taken with 2 different spheres, having a radius of 100 $\mu$m and 124 $\mu$m. The smaller one was used in a $^3$He cryostat, the other one in a dilution cryostat. The magnetic spheres were placed inside a horizontal parallel plate capacitor consisting of niobium electrodes to which a voltage was applied while the temperature was dropped below the superconducting transition temperature of niobium, ca.10 K. At helium temperatures the charged spheres were levitating between the capacitor plates. Frozen flux in the electrodes was necessary for a stable vertical oscillatory motion when a small ac voltage was applied at resonance. The frequency of the oscillations could not be controlled quantitatively because it depended on the frozen flux of the particular levitation status. It was observed, however, that a faster cooling rate from 10 K to 4.2 K in general lead to higher resonance frequencies. The critical velocities were measured by increasing the driving ac voltage from the linear flow regime to the turbulent one and back. The results of the various runs, all taken below 1 K, for both spheres are displayed in the Figure 1.   

\begin{figure}[ht]
\centerline{\includegraphics[width=\columnwidth,clip=true]{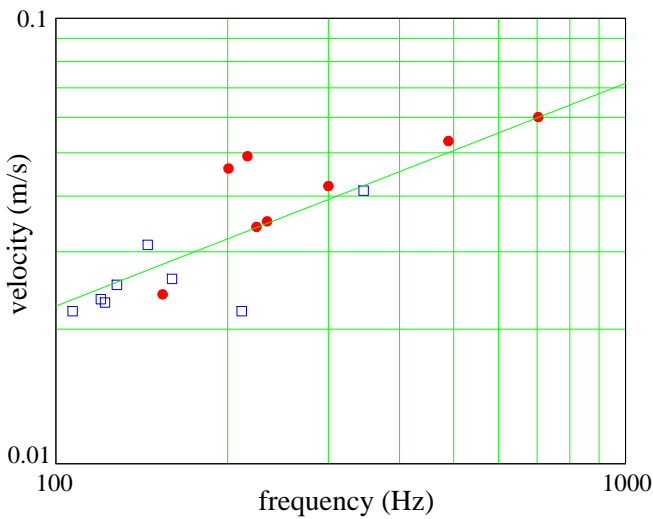}}
\caption{\label{f:vcsphere} Critical velocity for the onset of turbulence as a function of the frequency of 2 oscillating spheres. Blue squares: radius $124 \mu$m; red dots: radius $100 \mu$m; green solid line: $v_c=2.85\,\sqrt{\kappa 2\pi\!f}$ (m/s); temperature is below 1 K.}
\end{figure}

It is obvious that the predicted $\sqrt{\kappa\,\omega}$ variation gives a fair fit to the data. The errors of the individual values of $v_c$ vary from ca. 1\% to about 5\%, depending on how closely $v_c$ was approached in the experiments. The large scatter visible near 200 Hz is by far larger than any error bar. Recently, it has been found that remanent vorticity affects the critical velocity of vibrating wires and its absence greatly increased $v_c$ \cite{Yano}. This might be a reason for the scatter. Moreover, surface properties might affect a numerical prefactor in Eq.(\ref{e.scalerel}), although, as indicated above, only on some logarithmic scale, otherwise we would have had much more scatter. The mostly higher resonance frequencies of the smaller sphere are due to its smaller mass and probably also due the faster cooling rate from 10 K used with the $^3$He cryostat. Towards higher temperatures only a rather weak increase of $v_c$ by about 10\% was observed up to 2 K \cite{DissJ}. This rules out any thermally activated process for the onset of turbulence in these experiments. 

It will be interesting to compare these results for spheres with those for vibrating wires. Earlier observations by the Osaka group indicated frequency independent critical velocities below 2 kHz \cite{Yano1}. A more recent analysis by other authors \cite{Vinen} did show an increase scaling approximately with $\omega^{1/3}$. A more complete and very recent work of the Osaka group is in fact in agreement with Eq.(\ref{e.scalerel}) \cite{Yano2}.

We are very grateful to J. J\"ager, H. Kerscher, M. Niemetz, and B. Schuderer for their excellent thesis work in the former Low Temperature Group at Regensburg University, from which these data were obtained. W.S. has benefitted very much from his visits to the helium group lead by M. Krusius of the Low Temperature Laboratory at Helsinki University of Technology. R.H. acknowledges the support from the Academy of Finland (Grant 114887) and appreciates  discussions with M. Tsubota, H. Yano, M. Kobayashi, and S. Fujiyama. We both thank H. Yano very much for showing us his new results on the frequency dependence of the critical velocities of his vibrating wires before publication.

\section{Appendix 1}

One simple way to derive the scaling relation for $v_c$ (Eq.(\ref{e.scalerel})) is the following. The dispersion relation for Kelvin
waves with wave number $k$ is given by:
\begin{equation}
\omega(k) \approx \frac{\kappa k^2}{4\pi}\ln\left (\frac{2}{ka_0}\right).
\end{equation}
If we assume that the external drive at frequecy $\omega$ 
only drives the scales $R=2\pi/k$ given by the above dispersion relation, we find

\begin{equation}  
R \approx\sqrt{\kappa\pi\ln(2/ka_0)/\omega}.
\end{equation}
Inserting this then in Eq.~(\ref{e.critvel}) gives a critical velocity

\begin{equation}
v_c \approx \frac{\ln(16\pi/ka_0)}{2\pi\sqrt{\pi\ln(2/ka_0)}}\sqrt{\kappa\omega}.
\end{equation}
The coefficient in front should in principle be calculated self consistently using the dispersion 
relation but depends only very weakly on the frequency. For typical sphere experiments (200 Hz) the coefficient 
should have a value of 0.40 which is somewhat smaller than 2.85 obtained from the data of Fig.1. Of course, there is no real proof that 
Eq.(\ref{e.critvel}) should give the correct value in case of oscillating flows.

\section{Appendix 2}

The same dimensional arguments are applied now to a classical viscous liquid. For the stationary case the velocity scale is set by $\nu/R$, where $\nu$ is the kinematic viscosity. This is confirmed by comparing the laminar drag force of a sphere  $\lambda v$ with the turbulent drag force $\gamma v^2$. From Stokes' solution we have $\lambda=6\pi\nu\rho R$. Hence, by extrapolating both regimes, the drag forces are equal at a characteristic velocity 

\begin{equation}
v_{ch}=\frac{\lambda}{\gamma}=\frac{12\,\nu}{C_D\,R}.
\end{equation}

For the oscillating case we get from dimensional arguments a velocity scale $\sqrt{\nu\omega}$.
Stokes' solution for the linear drag is now \cite{Landau} $\lambda=6\pi\nu\rho R\cdot (R/\delta)$, where  $\delta=\sqrt{2\nu/\omega}$ is the viscous penetration depth, and it is assumed that $R\gg\delta$, which usually is the case for macroscopic bodies in liquid helium at or above the superfluid transition temperature. The velocity scale is now given by

\begin{equation}
v_{ch}= \frac{12}{\sqrt{2}\,C_D} \sqrt{\nu\omega}.
\end{equation}
Note that there is no length scale in this expression and, therefore, $v_{ch}$ does not depend on the size of the spheres.

Comparing these results with those for the superfluid, we note the similarity, except that the kinematic viscosity $\nu$ is replaced by the circulation quantum $\kappa$. The major differences, however, are, firstly, that in the classical liquid there are about three orders of magnitude in velocity between the Stokes regime of linear drag and fully developed turbulence and, secondly, because of the extrapolation, $v_{ch}$ is not a point on the $v(F)$ response of the body, whereas in the superfluid we have a true critical velocity of the oscillator for the onset of turbulent drag. Experiments with quartz tuning forks, for which analytical expressions of the drag forces are not available, confirm this classical scaling \cite{Ladik}.

%\end{multicols*} % ###multicol###

\end{document}